# ON THE CHALLENGES OF DATA PROVENANCE IN THE INTERNET OF THINGS

Mahmoud Elkhodr [1] and Zuhaib Bari Mufti[2]

[1] School of Engineering and Technology, Central Queensland University, Sydney, Australia
[2] School of Computing, Engineering, and Mathematics, Western Sydney University, Sydney, Australia

*ABSTRACT*

*The IoT is described as a smart interactive environment where devices communicate together ubiquitously sometimes in the background, performing functions on behalf of the users and offering many advanced services to them. Examples range from simple smart home applications such as ambient intelligence and remote controlling functionalities to more advanced smart cities setups. A smart IoT city for instance will encompass a network of many interconnected networks where various sensors and actuators distributed across many areas of the city share information, create knowledge and trigger actuation events. In such a dynamic and rich environment, it is vital for security to trace the source of data and verify its origin. This where data provenance in the IoT come to play. This work attempts to explore requirements and applications of data provenance in the IoT and the challenges pertaining to its realisation.*

## 1. INTRODUCTION

Earlier forms of provenance appeared as a method to validate the authenticity of an artefact by examining an object's origin, ownership or any modifications made to the item [2]. In a world entangled in a mesh of connected networks i.e. the Internet of Things (IoT), provenance becomes even more vital to keep track of events, the source of information, decisions, and origin of data and the metadata. E-Science relies on provenance to measure the quality of the data[1]. Nowadays, data provenance is no longer just concerned with finding the origin of the data, but it extends to include the capacity of tracking any events or modification made to the data. Example includes the followings applications[2]:

- Creating a file and any subsequent modifications to it and defining the ownership and accessibility is a form of File Systems provenance[1].
- Administrative systems and intrusion detection aided by logging system events is a form of Operating systems provenance.
- Similarly, compliers and run time errors can be detected by tagging the source line using compilers.
- Records of any insertion, modification and deletion are an application of provenance in curated databases[2].
- Browsing history is considered a form of web browsing provenance.

Additionally, several financial institutions are required by laws to record the source and origin of each digital transaction. This highlights the importance of provenance in the financial industry where each paper notes and its origin is treated as provenance. Intelligence and hospital systems are some of the prime users of provenance information [2]. A discrete information system having adequate relevance, capable of undergoing classification into various domains for the purpose of evaluation can be considered as Intelligence. Hospital records and related data protected by the Health Care Portability and Accountability Act (HIPAA) act makes it obligatory to record and store all hospital records and data in addition to managing proper authorised access to the data[3]. Information and lineage data used as provenance must possess some inherent technical features in order for it to be reliable. Some of these features are as follows:

- Information about every action performed on data needs to be preserved and stored completely[4].
- Ensuring that no manipulation of the data with a malicious intent takes place (Integrity).
- Provenance data should be available readily without any hassle (Availability).
- By providing authorised access to provenance data, confidentiality of the information can be ensured[5].
- Provenance data in the E-science field must be obtained in an economically feasible manner.
- Provenance data must be stored and available in such a way that the privacy of a person is not compromised, especially in the IoT[6]. Systems involving data provenance data need to deal with diverging aspects of ensuring that no outside entity or system is able to access the data and at the same time data within the system is readily available and shared among authorised entities for transparency[5].

## 2. APPLICATIONS OF DATA PROVENANCE

Some of the most common applications of provenance have been listed below:

**DIAGNOSTICS**: Provenance has been used for debugging and detecting real time anomalies in a distributed system[7]. If a monitoring system is based on declarative monitoring, there is a provision to analyse the network traffic which indirectly can be employed for detecting an intrusion[8]. SeNDlog can dynamically trace changes to a routing table and helps in generation of an alarm if the number of changes made are above a certain threshold value. Once an alarm has been generated a distributed recursive query on the network performance can trace the origin of any malicious activity[9].

**SECURITY**: Data provenance covers historical data in addition to real time data as well. This helps in finding correlations in the network pattern of an attacker; thus, helping in the security of vital assets. Locating the source or filtering the IP address from the traffic is a typical example[10]. Annotations can be used in data provenance to help identifying potential attacker as well as tracing back information for forensic analysis[8]. Provenance can also be used to identify any malicious packets dropping in a sensor network[11].

**ACCOUNTABILITY:** Data Provenance ensures a proper accountability for an action as well as data. In conventional forensic analysis, call-details consisting of information, time and location of the call are a form of data provenance. Network Provenance can be also used to manage trusts in a distributed environment[12].

**TRUST**: By enabling a network of information where nodes are capable of tracing the origin of data, effective trust policies can be implemented[8]. Multi-hop networks and Body Sensor Networks rely also on data provenance to ensure trust[13]. Provenance can be used in quantifying trust, which enables sensors to process information from trusted nodes only (Wenchao et al., 2008).

**OPTIMIZATION**: Monitoring of a system and tracing important events using data provenance in sensor networks can help in optimization of resources[2]. Resource allocation and finding bad routes or draining nodes are good examples.

**DEVELOPMENT PROCESS**: Provenance logs can be used to capture changes in a network before and after an event takes place[14]. Comparing the snapshots before and after to see the changes in energy and other resources and using provenance to gauge the dependencies of a system can help in the development of a smooth process.

**RECOVERY**: Provenance is often used to restore a system after a failure and for success validation[2]. In a sensor network, it is vital to not only identify the points of failure but also to avoid those which cause system anomalies. Provenance of graphs plays a key role in scenarios requiring troubleshooting as well.

## 3. DATA PROVENANCE CHALLENGES IN THE IOT

The IoT proposes various revolutionary concepts by employing millions, even billions, of tiny sensor or actuators nodes collecting and communicating information just about everything[15]. The volume of data collected in such a large network will have a high velocity, volume and divergent variety. This augments the significance of analysing the data for trustworthiness establishment in order to make better decisions. Therefore, it is becoming increasingly important to analyse a distributed network for possible anomalies and to pinpoint any erring node. These capabilities are some of the functional requirements needed to provision for Network Accountability and forensic analysis. Therefore, provenance of information or data plays a critical role in such environments. On the other hand, in an IoT smart based environment, the flow of information is relayed ultimately through the open Internet. It is a well-known security principle that the Internet is insecure. Therefore, it is essential to have reliability, trust, accountability and similar security principles addressed by employing a strong provenance enabled system.

To this end, as new, complex and dynamic data exchanged by IoT devices gets published on the Internet -where platforms accessing, publishing and modifying the data can be also diverse-, it becomes important to address the lineage, trustworthiness, reliability and accuracy of data in the IoT[16]. While papers' provenance has been employed in several systems, the IoT poses some unique challenges to the provisioning of data. Some of the challenges are listed below.

**SECURITY**

Data transmitted through an IoT system is extremely susceptible to attacks by a third party[17]. If provenance of data is insecure, it can result in a breach of sensitive information. The challenge is to impart enough confidentiality so that provenance can be accessed by only authorised individuals. Under certain circumstances, identity and location of the IoT device needs to be secured above all as the device may be more valuable than the data it sends. A robust security mechanism should incorporate confidentiality, integrity, privacy and availability of the information[18]. However, a high level of heterogeneity coupled with the massive scale in which IoT devices are likely to be deployed complicates the security issue of data provenance in the IoT[18]. Moreover, IoT devices lack the computational power and energy requirements to incorporate complex security solutions such as encryption, cryptography, public key and symmetric key infrastructure[19]. Integrity of data provenance to assure a level of trust should be considered as well. This demands the use of cryptographic hashes algorithms which are extremely difficult to implement in the IoT due to the resource constraint feature of IoT devices[17].

**BIG DATA**

The massive volume of data produced by sensor networks in the IoT can result in the generation of petabytes of data, thus resulting in additional computational burden on the already fragile system [20]. Some researchers point out to the fact that Big Data and IoT need to be treated in tandem rather than as separate entities [21]. Querying and tracing Provenance information in such a system to point out the anomalies and other faults in the system is extremely difficult. Data Provenance may consume a lot of network resources, which in turn may hamper the operational efficiency of the system [17]. To ensure that Metadata is readily available upon request, there is a need to design systems which have a very low computational overhead to ensure smooth performance [22].

**INDEXING:**

A complete list of provenances in an IoT environment is practically impossible owing to the large nature of information. Hence, an indexing scheme is normally used [22]. However, it is likely that information can't be queried in a conventional manner wherein looking-up an attribute to retrieve the data is common. Users often must query the dataset, which is essentially a subset of an attribute. Even in XML-based schema used for mapping names and values may prove not to be enough without the help of additional structures.

**MULTIPLE CONSUMERS**:

IoT data can have potentially vast and diverse range of consumers, with clients possessing divergent requirements. Some clients may need data on a real time bases, whilst others may just need to archive the provenance data. For example, while managing a smart city environment, provenance data may be required dynamically to make better decisions and rectify any anomalies in a system. Therefore, adequate flexibility is required for the provenance of data in the IoT.

**TRANSFORMATION OF DATA:**

Sensors in an IoT network collect data and pass or route them to other sensors, which may modify the information before passing it on to a more computationally powerful device. In other cases, actuators may receive data modified by various sensors during the transfer phase and thus, it becomes necessary to overcome the challenges encountered in representing such a complex provenance of information [17].

**QUERYING INFORMATION:**

Just tracing the lineage of data and its object may not be adequate for future systems and powerful querying tools need to be deployed to meet the cybersecurity challenges of next generation [23]. Records may need to be queried based on the context and requirements while maintaining the confidentiality at the same time may be essential [17].

**INTEROPERABILITY:**

IoT devices need to work in an extremely interoperable environment to ensure that the data collected by the sensors is successfully delivered to the target location. Also, various intermediate nodes or platforms are capable of reading or modifying the data. In such as case, Data Provenance demands that all the devices present in a system to be interoperable by having sufficient features to use each other's data. Keeping in view the limited computational power and resources of IoT devices and ensuring security of the system, achieving efficient interoperability in the IoT is still not an easy task[24]. IoT devices are manufactured by different vendors and may use different networking and routing protocols and often there is no standard or regulation yet in place to ensure uniformity and interoperability of devices.

**DATABASE MANAGEMENT:**

IoT data can be discrete, continuous, and dynamic. Certain data can be descriptive or based on environmental factors. Other can be in the form of addresses such as RFID tag format [25]. As the number of IoT devices may run into Billion coupled with limited computational capability of devices, it is almost impossible to adhere to IPv4 protocol for IoT Devices. Thus Internet Engineering Task Force (IETF) has introduced various protocols for IoT based in IPv6 addressing format [26]. But in doing so the header size has been increased from 32 bit to 128 bit addressing scheme, thus making it extremely difficult for resource constrained IoT devices to implement the system [25]. Thus, traditional databases may not provide a complete solution for such a complex system and it becomes imperative to deploy innovative and non-traditional databases.

An innovative approach is needed to cope up with the challenges associated with data provenance in IoT. In this case numerous protocols have been put forward such as the 6LoWPAN (IPv6 over low power wireless personal area network) protocol which is specially designed for resource constrained devices. The protocol is based on IPv6 and ensures universality, stability and additional features for IoT devices [26]. 6LoWPAN protocol suite specifically targets the integration of IPv6 and MAC (Media Access Control) and physical layers used in IEEE 802.15.4 standard. It is pertinent to mention that the maximum frame size of 127 bytes supported by IEEE 802.15.4 standard hinders the use of IPv6 and MAC header. By incorporating such a technology, it is possible to address various security and provenance issues using symmetric key and public key cryptography solutions.

One must also consider that not all IoT devices can transmit data. Hence, IoT gateways are used in some cases to bridge between the IoT devices with the Internet. Therefore, helping in harnessing the full potential of the technology [27]. The gateways provide a mechanism to ensure the computational power of IoT devices does not need to be high enough to increase the overall cost of the system, but at the same time they are able to smoothly operate in tandem with external applications and computational devices without compromising the efficiency and effectiveness of the system. Constricted application Protocol (CoAP) for device to device communication is employed to enables IoT devices to use the Representational state transfer (REST) mechanism which is similar to HTTP. This enables data provenance to be written using standard HTTP queries, which helps in mitigating the complexities of collecting provenance of data in IoT applications. The use of several NoSQL like CouchDB, MongoDB etc. databases to store provenance data is recommended as they enable extensive flexibility during storing and retrieving of information.

## 4. PROVENANCE IN IoT SENSOR NETWORKS

Sensor networks are increasingly being deployed in battlefield and in other critical areas demanding impeccable level of provenance data [15]. A novel approach of using the trust scores of provenance data can help to quantify the trustworthiness of provenance information [15]. Nodes that are a part of the sensor network as well as the data, both have their trustworthiness quantified in terms of trust scores [15]. The approach relies on a cyclic framework wherein the trust score of provenance data impacts the trust score of a sensor node and vice versa, thus showing a strong interdependent relationship between the two properties.

Each data item possesses two critical properties. i.e. value similarity and provenance similarity. In case an event has higher number of similar values obtained from different nodes, the trust score of the value is deemed to be high while as diverse nature of data provenances having similar values signifies a higher provenance similarity [15]. In a sensor network consisting of many nodes, a server node is tasked with the collection of all the sensors' data passed on to it by a server node. Data generated by a server node gets passed to the intermediated nodes, which either modifies or directly passes it to another intermediate node or the sensor node depending upon the proximity of the node. Provenance in this case is defined as how and where the data was generated and the steps involved during transferring of data from terminal node to server node via intermediate nodes [15]. It can also be regarded as the sub-graph of the network with nodes further classified as simple nodes having only one child and associated with transferring of data from one node to another. Simple nodes are an integral part of an ad-hoc sensor network tasked with relaying the data to server in order to cope up with lack of network infrastructure [15]. Certain internal nodes receive data items from more than one node. They then process the data into a single entity before passing it to the server node acting as aggregate nodes [15].

Sensor data collected in an IoT environment needs to be combined with data obtained from other sensors in addition to being location specific [2]. For instance, the data obtained from sensors which monitor traffic in a smart city can help in ticketing the offenders but when combined with sensor data from weather monitoring systems and from other cities, then a statistical observation can be deduced. Hence, it can prove to be extremely beneficial to manage provenance data effectively and efficiently. Internet of Things consisting of a vast network of sensing devices coupled with actuators mitigates the necessity to have a proper Provenance model in tandem with a robust management of provenance to overcome the trust, pedigree and security challenges associated with it. Any provenance management system has to deal with three fundamental issues:

How to capture the data?
   1. Where and how to store the data?
   2. How to query the data?

Data Provenance can be captured as a workflow, as a process or using an operating system [2]. Unlike other systems, workflows support provenance from the design phase and hence have a better model for provenance, while as a process-based system deduces provenance using information extracted from various processes. In case of an operating system based provenance, it becomes necessary to opt for post processing mechanism to gain provenance information as the workflows aren't integrated with them [28]. Provenance information often raises an important question pertaining to the granularity of data. In sensor networks provenance can either be stored in a coarse grained or fine grained manner. In case of a coarse grained provenance, dependencies between the data items are tracked on an abstract level, giving an overview of the general process rather than delving into the detailed information regarding each individual item as done in a fine grained provenance model [29]. It is advisable to implement fine grained provenance to track the source of data and any subsequent nodes performing modifications to the data. This is useful in situations like monitoring the temperature and pressure in an industrial set up. This is important for human operators to know the exact location and point of concern. In republishing system and applications that require obtaining a streaming of senor data, fine grained applications are preferred over coarse grained ones [30], [31].

Provenance capturing mechanisms can be automated or manual depending upon the field of applications and the IoT devices used. In automated monitoring or observed provenance systems, the relevant data is stored/gathered automatically without relying on a manual input [2]. Parameter selection for a monitoring process can be enabled by an administrator without directly interfering in the process of provenance data collection in an automated monitoring fashion. An automated monitoring system may be preferred in certain circumstances as it is more resistant to malicious attacks primarily because users doesn't have the authority to alter the provenance data. However, owing to the extreme heterogeneity of IoT systems, it becomes extremely cumbersome to integrate the provenance data with other applications without the aid of designated software. In disclosed or manual provenance system, sensors rely on input from a user. Thus, these data are prone to manipulation by malicious entities. However, the integration of a manual capturing process is far more convenient to use than the use of automated capturing systems as they do not require the aid of a software [2].

Once provenance data has been collected in an IoT system, it is necessary to store these data somewhere. This storage issue poses an operational challenge in the IoT. In case local provenance is used in an IoT system, some studies such as [9] recommended storing the whole data including a reachability tree at the local node. In distributed provenance systems, it is advisable to store the pointer to previous nodes in order to reconstruct the provenance needed on a demand basis [9]. Provenance can also be used to store the state of a network at any given point of time to detect run time anomalies. This can be considered as an example of online provenance applications. Offline provenance can be kept even after objectives have been met, primarily to keep historical data in order to develop the reachability tree between the network nodes [2]. However, such a system will have a high redundancy and storage overhead; thus, creating economical and resource issues. Provenance data once collected can be stored at a single centralised location. However, despite having a direct connection between the data and its meta-data, there are difficulties in maintaining provenance due to the heterogeneous nature of the IoT devices and the data they exchange. When data are stored in a distributed environment, maintenance becomes far more convenient, but querying of data is much more complicated than that used in a centralised environment [32]. As provenance data can grow exponentially resulting in a possible scenario wherein the provenance data are far larger than the actual data in terms of storage value. By recursively traversing the source of data and preferring inversion method over annotations method to describe an attribute, the magnitude of storing provenance data can be vastly reduced [9]. Data redundancy can be reduced in addition to improving network storage utilization in provenance systems by employing deduplication schemes [33]. Deduplication works by relegating multiple records of the same provenance chain as many objects in a wireless sensor network may have similar ancestors [33]. In case of a geographical provenance, using city-based provenance over a state level provenance can help to achieve a higher degree of scalability [15].

Querying Provenance data in an IoT is not a straightforward task when compared to querying provenance data in scientific workflows and web as data captured by a node may be a precursor to an actuator or may even be used by the same node [34]. Provenance data is created at a far greater rate than it is ever queried, an undeniable fact of provenance data (Dogan, 2016a). Without an efficient querying mechanism, the value for provenance data may get severely degraded [34]. Provenance data may be in the form of a graph and as such query languages must be advanced enough to interoperate with graphs. For example, a reverse traversal of Directed

International Journal of Wireless & Mobile Networks (IJWMN) Vol. 11, No. 3, June 2019Acyclic Graph (DAG) when queried may produce an output which is a subset of the original provenance DAG [35]. Owing to the user knowledge and acceptance of languages like SQL, Prolog, which lacks essential features required for a data provenance language, may still be preferred by certain users [2].

## 5. CONCLUSIONS

This paper discussed the issue of achieving data provenance in the IoT. It considers Data Provenance as one of the pillars requirements in the IoT. In a complex, dynamic, and heterogenous interconnected system such as that encountered in the IoT, it is vital to determine the source and origin of data. Not only that, it is also of equal importance to be able to identify the players or processes that contributed to changes in data. This is a necessary in order to determine whether the data supplied by an IoT device or a system can be treated as trustworthy. Consequently, this paper presented some of the main requirements of data provenance in the IoT, the envisioned applications, and, importantly, the challenges pertaining to data provenance in the IoT. To this end, this review clearly points to the need to further explore data provenance solutions in the IoT. Such solutions need to not only provide information on the source of the data or the events, devices, applications, processes that contributed to changes of data, but also the solution itself need to be trustworthy, secure, and light in terms of computational resources. This opens the door to many additional challenges such as storage of data provenance, access rights and privileges, privacy, among others that call on further research in this area.